\documentclass[showpacs,preprintnumbers,prd,nofootinbib,floats,amssymb,floatfix]{revtex4}
\usepackage{amsmath}
\usepackage{amsfonts}
\usepackage{graphicx}
\usepackage{newtxtext}
\usepackage{newtxmath}
\usepackage{euler}
\usepackage{amsxtra}
\usepackage{eufrak}
\usepackage{hyperref}
\usepackage{amsmath}
\usepackage{amstext}
\usepackage{endnotes}
 
\begin{document}
\title{ Constrained dynamics of  two interacting relativistic particles in the Faddeev-Jackiw symplectic framework}
\author{ Omar Rodr{\'i}guez-Tzompantzi} \email{omar.tz2701@gmail.com}
\affiliation{ Instituto de Ciencias Nucleares, Universidad Nacional Aut\'onoma de M\'exico,
Apartado Postal 70-543, Ciudad de M\'exico 04510, M\'exico.}
\begin{abstract}
The Faddeev-Jackiw symplectic formalism for constrained systems is applied to analyze the dynamical content of a model describing two massive relativistic particles with interaction, which can also be interpreted as a bigravity model in one dimension. We systematically investigate the nature of the physical constraints, for which we also determine the zero-modes structure of the corresponding symplectic matrix. After identifying the whole set of constraints, we find out the transformation laws for all the set of dynamical variables corresponding to gauge symmetries, encoded in the remaining zero-modes. In addition, we use an appropriate gauge-fixing procedure, the conformal gauge, to compute the quantization brackets (Faddeev-Jackiw brackets) and also obtain the number of physical degree of freedom. Finally, we argue that this symplectic approach can be helpful for assessing physical constraints and understanding the gauge structure of theories of interacting spin-2 fields.
\end{abstract}
 \date{\today}
%\pacs{}
\preprint{}
\maketitle
\section{Introduction}
In recent years the study of alternative gravity theories beyond Einstein's general relativity,  such as massive gravity, bi-gravity and multi-gravity theories, has long been of considerable interest, both theoretically and  phenomenologically \cite{Rham1, Rham3, aspects, interacting, phenomenology, ac-uni, ac-uni2, dark,Infraded, black-holes, massive, Hassan0,Hassan1,Hassan}. The construction of this kind of theories must ensure the absence of any unphysical degrees of freedom - the Boulware-Deser ghost, thereby rendering a stable and consistent theory \cite{ghost}. At the Lagrangian/Hamiltonian level,  in order to avoid this ghost degree of freedom, the ghost-free theories must possess the necessary physical constraints for getting rid of both the ghost field and its canonically conjugate moment. The standard procedure for determining these constraints is Dirac's formalism of constrained Hamiltonian systems (see e.g. \cite{ Hassan, blagojevic,Kluson}). In this approach, the primary constraints appear when the canonical momenta are computed. Consistency of the theory requires that the primary constraints remain on the constraint surface during their evolution. These consistency conditions will lead to the secondary constraints, and so on. Thereafter, the whole set of constraints must be classified into first- and second-class ones, and then a generator of the local gauge transformations can be constructed as a suitable combination of the first-class constraints. The number of physical degrees of freedom can be also explicitly counted, and a kind the brackets (Dirac's brackets) to quantize gauge systems can be obtained \cite{Dirac, teitelboim, Henneaux, omar2}. Nevertheless, Dirac's analysis of these models  can become technically intricate and obscure. More specifically, it could be difficult to identify the whole set of physical constraints, which means that the classification and separation of all the constraints into first- and second-class ones can become non-trivial, which can hide the dynamical structure of these theories. Subsequently, the important task of understanding the symmetry properties of this kind of theories by considering the first-class constraints has not been discussed in the literature yet, and the Dirac's bracket associated to the second-class constraints was not computed either.

In a recent paper \cite{dominici}, in order to have a better understanding of how to deal with this kind of dynamical systems, a simple model  was considered  that describes two relativistic particles with interaction \cite{G1,G2},  which possesses many interesting properties since this model can be interpreted  as a bigravity model in one dimension. The primary, secondary, tertiary and quaternary constraints were derived \cite{dominici} by using Dirac's Hamiltonian formalism. However, it turns out that the procedure for separating the second-class from the first-class constraints is not completely clear. In addition,  the gauge transformations and their generator are not shown, and the quantization brackets associated with the cancellation of the second-class constraints were not computed either, which result in an incomplete and rather complex Dirac's Hamiltonian analysis. These arguments motivate the study of the dynamical structure for this kind of theories from another point of view.

The aim of the present paper is to study this problem in an alternative approach from the one presented in Ref. \cite{dominici}.  In particular, we will apply the symplectic approach developed by  Faddeev and Jackiw \cite{Faddeev}. The advantage of the symplectic approach is that we do not need to classify the constraints into first- and second-class ones. Even more, the Faddeev-Jackiw formalism does not rely on Dirac's conjecture.  Additionally,  in this method, there are less constraints as compared with those generated by the Dirac approach. Therefore, we can expect that the algebraic manipulations needed in the treatment of the constrained systems could be shortened. In particular, if secondary, tertiary, or higher-order constraints are present.  Nevertheless, in Ref. \cite{Antonio} it has been shown that, though technically different, the Faddeev-Jackiw approach and Dirac method are physically equivalent. A recent application of Faddeev-Jackiw (F-J) formalism was discussed in \cite{omar}, where topologically massive gravity was studied ( other interesting works can be found in references cited here).  As will be shown below, in this framework, the nature of the constraints, the gauge symmetries (and their corresponding generators) and the quantization brackets can be systematically addressed by investigating the properties of the symplectic two-form matrix and its corresponding zero-modes \cite{barcelos, barcelos2, Montani1, Montani2}.

The structure of this paper is as follows. In Sec. II  we show that the F-J symplectic method applied to two interacting  relativistic particles allow us in a simple way to identify the complete set of dynamical constraints. In Sec. III, we show that the gauge transformations can be computed using the zero-modes of the symplectic 2-form matrix. Furthermore, using the gauge transformations we show that the Lagrangian is invariant under these gauge symmetries.  In Sec. IV, we determine both the fundamental F-J brackets and the physical degrees of freedom by introducing a gauge-fixing procedure. In Sec. V, we present a summary and the conclusions.
%%%%%%%%%%%%%%%%%%%%%%%%%%%%%%%%%%%%%%%%%%%%%%%
\section{Constraint structure in the Faddeev-Jackiw framework} 
In a given inertial frame of reference, the system of two relativistic particles will be described by two word-lines given in terms of the space-time coordinates $x_{i}^{\mu} (i=1,2)$. The configuration space of the system is the product of two Minkowski spaces, $\mathcal{M}\times\mathcal{ M}$. In this space, the world-lines $\gamma$ and $\gamma'$ of the two particles will correspond to a two-dimensional surface $\gamma\times\gamma'$. In this way, in order to introduce an interaction between the particles, one can merely postulate the existence of a one-to-one correspondence of between the events of the two world-lines. This correlation means choosing a given line $l$ on $\mathcal{M}\times\mathcal{M}$ among the infinite ones lying on $\gamma\times\gamma'$, which connect the initial and the final pairs of events. Additionally, if we incorporate an invariant   parameter  $\tau$ in order to parametrize the line $l$, then the equations of motion will be ordinary differential equations in this unique parameter \cite{dominici,G1,G2}. 

In Ref. \cite{dominici}, the model of two massive relativistic particles interacting  is described by the following action\footnote{the signature of the metric is $\left( -,+,+,+\right)$.}
\begin{equation}
S[x_{1}, x_{2},e_{1},e_{2}]= \int d\tau\mathcal{L}=\int d\tau\left[\frac{\dot{x}_{1}^{2}}{2e_{1}}+\frac{\dot{x}_{2}^{2}}{2e_{2}}+\frac{e_{1}}{2}\left(m_{1}{^{2}}-k^{2}r^{2}\right)+\frac{e_{2}}{2}\left(m_{2}{^{2}}-k^{2}r^{2}\right)\right],
 \label{eq:p9}
\end{equation}
where $r^{2}=r\cdot r=\eta^{\mu\nu}r_{\mu}r_{\nu}$ $(\mu,\nu=0,1,2,3)$ and $r^{\mu}=\left(x_{2}^{\mu}-x_{1}^{\mu}\right)$ is the relative distance, with $x_{1}^{\mu}$ and $x_{2}^{\mu}$  the space-time coordinates for the two particles, and  $m_{1}$ and $m_{2}$ represent the rest masses of the particles, respectively.  Furthermore,  $e_{1}$ and $e_{2}$ are Lagrange multipliers that can be interpreted as two einbein in Polyakov's formulation. All these variables are function of the evolution parameter $\tau$, therefore the dot stands for a derivative with respect to $\tau$. To proceed with the analysis of the constraints  within Faddeev-Jackiw symplectic framework \cite{Faddeev, barcelos, barcelos2, Montani1, Montani2}, one needs to convert the Lagrangian (\ref{eq:p9}) into a first-order Lagrangian. For this porpuse,  one can define the canonical momentum $( p_{1}^{\mu}, p_{2}^{\mu},\pi_{1},\pi_{2})$  associated to $(x_{1}^{\mu},x_{2}^{\mu},e_{1},e_{2})$ respectively, as
\begin{equation}
p_{1}^{\mu}\equiv\frac{\partial{\mathcal{L}}}{\partial \dot{x}_{1\mu}}=\frac{\dot{x}_{1}^{\mu}}{e_{1}},\quad p_{2}^{\mu}\equiv\frac{\partial{\mathcal{L}}}{\partial \dot{x}_{2\mu}}=\frac{\dot{x}_{2}^{\mu}}{e_{2}},\quad{\pi}_{1}\equiv\frac{\partial{\mathcal{L}}}{\partial\dot{e}_{1}}=0,\quad{\pi}_{2}\equiv\frac{\partial{\mathcal{L}}}{\partial\dot{e}_{2}}=0.\label{momentum}
\end{equation}
The Lagrangian (\ref{eq:p9}) can thus be rewritten as follows:
\begin{eqnarray}
{\mathcal{L}}=p_{1\mu}\dot{x}_{1}^{\mu}+p_{2\mu}\dot{x}_{2}^{\mu}-\frac{1}{2}e_{1}\left(p_{1}{^{2}}-\left(m_{1}{^{2}}-k^{2}r^{2}\right)\right)-\frac{1}{2}e_{2}\left(p_{2}{^{2}}-\left(m_{2}{^{2}}-k^{2}r^{2}\right)\right).\label{hamiltonian}
\end{eqnarray}
This is the form of the Lagrangian that we will study from now on.

 The variation with respect to the dynamical variables $x _{1}^{\mu}$, $p _{1}^{\mu}$, $x _{2}^{\mu}$, $p _{2}^{\mu}$, $e _{1}$ and $e _{2}$, yields the following set of equations of motion:
\begin{eqnarray}
\left(\delta x_{1}^{\mu}\right)&:&\dot{p}_{1}^{\mu}-k^{2}\left(e_{1}+e_{2}\right)r^{\mu}=0,\label{eom1}\\
\left(\delta p_{1}^{\mu}\right)&:&\dot{x}_{1}^{\mu}-e_{1}p_{1}^{\mu}=0,\\\label{eom2}
\left(\delta x_{2}^{\mu}\right)&:&\dot{p}_{2}^{\mu}+k^{2}\left(e_{1}+e_{2}\right)r^{\mu}=0,\\\label{eom3}
\left(\delta p_{2}^{\mu}\right)&:&\dot{x}_{2}^{\mu}-e_{2}p_{2}^{\mu}=0,\\\label{eom4}
\left(\delta e_{1}\right)&:&\frac{1}{2}\left(p_{1}{^{2}}-\left(m_{1}{^{2}}-k^{2}r^{2}\right)\right)=0,\\\label{eom5}
\left(\delta e_{2}\right)&:&\frac{1}{2}\left(p_{2}{^{2}}-\left(m_{2}{^{2}}-k^{2}r^{2}\right)\right)=0\label{eom6}.
\end{eqnarray}
Combining (\ref{eom1}) and (\ref{eom2}), one  can deduce that the total momentum defined as $P^{\mu}\equiv\left(p_{2}^{\mu}+p_{1}^{\mu}\right)$ is conserved.
Notice also that the above Lagrangian (\ref{hamiltonian}) can be expressed simply in the symplectic form,
\begin{equation}
{\mathcal{L}}(\xi, \dot{\xi})=a_{I}(\xi){}\dot{\xi}^{I}-V(\xi),\label{LS}
\end{equation}
where the symplectic potential $V$ in Eq. (\ref{LS}) is given by
\begin{equation}
V(\xi)=\frac{1}{2}e_{1}\left(p_{1}{^{2}}-\left(m_{1}{^{2}}-k^{2}r^{2}\right)\right)+\frac{1}{2}e_{2}\left(p_{2}{^{2}}-\left(m_{2}{^{2}}-k^{2}r^{2}\right)\right).
\end{equation}
Furthermore ${\xi}{^{I}}= (x_{1}^{\mu}, p_{1}^{\mu}, x_{2}^{\mu}, p_{2}^{\mu}, e_{1}, e_{2})$ and ${a}{_{I}} = ( p_{1\mu}, 0, p_{2\mu}, 0, 0, 0 )$ stand for the initial set of symplectic variables and the corresponding canonical one-form, respectively. As shown in Appendix A, the corresponding symplectic equations of motion that arise from the Lagrangian (\ref{LS}) can be expressed as 
\begin{equation}
f_{IJ}\dot{\xi}^{J}-\frac{\partial }{\partial\xi^{I}}V(\xi)=0.\label{eqmot}
\end{equation}
The dynamics of the model is then  characterized by the symplectic two-form matrix, $f_{IJ}\equiv\frac{\partial }{\partial\xi^{I}}a_{J}-\frac{\partial }{\partial\xi^{J}}a_{I}$, which is antisymmetric. In our case, the symplectic matrix takes the form 
\begin{eqnarray}
\label{eq13}f{_{IJ}}=
\left(
  \begin{array}{cccccc}
 0   & -\eta^{\mu}_{\nu}   &  0   &  0     &  0  &  0 	 	 \\                                                                        
 \eta^{\mu}_{\nu} & 0   & 0   &  0 &   0   & 0 \\                                                                   
0   &  0   &  0      &  -\eta^{\mu}_{\nu}   &   0   &  0 	\\
    0   & 0   &\eta^{\mu}_{\nu}    & 0  & 0	 &  0 \\
0  & 0  &  0 &0  & 0 	& 0 	 \\
0   &  0  & 0  & 0   &  0 	&  0 
 \end{array}
\right),
\end{eqnarray}
which clearly is not invertible because, det $\parallel f{_{IJ}}\parallel =0$. This means that we have a constrained theory with a kernel of dimension $2$, generated by the  following zero-modes $\left(v{_{1}}\right)_{I}= \left (0,0 ,0,0, v^{e_{1}},0 \right)$ and $\left(v{_{2}}\right)_{I}= \left(0,0,0,0,0,v^{e{_{2}}}\right)$ such that $\left(v_{1,2}\right)_{I}f_{IJ}=0$; here $v^{e{_{1}}}$ and $v^{e{_{2}}}$ are arbitrary functions. According to the F-J procedure, the multiplication of the  equations of motion (\ref{eqmot}) by the zero-modes leads to the following constraint equations:
\begin{eqnarray}
 \left(v{_{1}}\right)_{I}\frac{\partial}{\partial\xi^{I}}V(\xi) &=&v^{e_{1}}\frac{1}{2}\left(p_{1}{^{2}}-\left(m_{1}{^{2}}-k^{2}r^{2}\right)\right)=0,\label{p1} \\
\left(v{_{2}}\right)_{I}\frac{\partial}{\partial \xi^{I}}V(\xi)&=&v^{e_{2}}\frac{1}{2}\left(p_{2}{^{2}}-\left(m_{2}{^{2}}-k^{2}r^{2}\right)\right)=0.\label{p2}
\end{eqnarray}
Since $v^{e_{1}}$ and $v^{e_{2}}$ are any arbitrary functions, we get the  following constraints:
\begin{eqnarray}
\Phi_{1}&=&\frac{1}{2}\left(p_{1}{^{2}}-\left(m_{1}{^{2}}-k^{2}r^{2}\right)\right)=0,\label{pry1}\\
\Phi_{2}&=&\frac{1}{2}\left(p_{2}{^{2}}-\left(m_{2}{^{2}}-k^{2}r^{2}\right)\right)=0,\label{pry2}
\end{eqnarray}
which are a modification of the mass shell conditions for  two free particles through the interaction term (potential term).
 
On the other hand, with the aim of deducing new constraints, one can introduce the consistency condition (time conservation) on the constraints (\ref{pry1}) and (\ref{pry2}) which guarantees that the constraints remain on the constraint surface during their evolution. Consequently, the consistency condition for the constraints (\ref{pry1}) and (\ref{pry2}) reads
\begin{equation}
\dot{\Xi}=\frac{\partial \Xi}{\partial \xi^{I}}\dot{\xi^{I}}=0 \hphantom{111}\mathrm{with}\hphantom{111} \Xi=\left(\Phi_{1},\Phi_{2}\right).\label{consis}
\end{equation}
These conditions can be incorporated into the equation of motion (\ref{eqmot}). In this
way, one has the following new linear equations:
\begin{equation}
f^{(1)}_{KJ}\dot{\xi}^{J}= Z^{(1)}_{K}(\xi),\label{equation2}
\end{equation}
with
\begin{equation}
f^{(1)}_{KJ}=
\left(
\begin{array}{cc} 
f_{IJ} \\ 
\frac{\partial\Xi}{\partial\xi^{J}}
\end{array}
\right)\hphantom{111}\&\hphantom{111}Z^{(1)}_{K}(\xi)=
\left(
\begin{array}{ccc} 
\frac{\partial V}{\partial \xi^{I}} \\ 
0 \\
0\\ 
\end{array}\label{equation2}
\right).
\end{equation}
The matrix $f_{IJ}$ of (\ref{equation2}), is exactly the  expression given in Eq. (\ref{eq13}), and the matrix $\frac{\partial \Xi}{\partial\xi^{J}}$ simply reads
\begin{equation}
\label{eq}
\frac{\partial \Xi}{\partial\xi^{J}}=
\left(
  \begin{array}{cccccc}
-k^{2}r^{\mu}   & p_{1}^{\mu} & k^{2}r^{\mu}  & 0 &0 &0  \\
-k^{2}r^{\mu}  & 0 & k^{2}r^{\mu} &p_{2}^{\mu} & 0 & 0 
 \end{array}
\right),
\end{equation}
whereas the rectangular matrix $Z_{K}^{(1)}$ becomes:
\begin{equation}
\label{z1}
Z_{K}^{(1)}=
\small{
\left(
  \begin{array}{c}
-k^{2}\left(e_{1}+e_{2}\right)r_{\nu} \\
e_{1}p_{1\nu} \\ 
k^{2}\left(e_{1}+e_{2}\right)r_{\nu} \\
e_{2}p_{2\nu}\\
\Phi_{1} \\
\Phi_{2} \\
0 \\
0
 \end{array}
\right).}
\end{equation}
It is easy to see that the matrix $f_{KJ}^{(1)}$ is degenerated, and its zero-modes are
\begin{eqnarray}
\left(v^{(1)}{_{1}}\right)_{K}&=& \left(p_{1}^{\nu}, k^{2}r^{\nu}, 0, -k^{2}r^{\nu}, 0, 0, 1, 0 \right),\label{V4}\\
\left(v^{(1)}{_{2}}\right)_{K}&=& \left(0, k^{2}r^{\nu}, p_{2}^{\nu}, -k^{2}r^{\nu}, 0, 0, 0, 1 \right)\label{V3}.
\end{eqnarray} 
As before, the new constraint should be obtained multiplying the Eq. (\ref{z1})  by the zero-modes (\ref{V4}) and (\ref{V3}) respectively, i.e.,
\begin{eqnarray}
\left(v{^{(1)}}_{2}\right)_{K}Z^{(1)}_{K}|_{\Xi=0}&=&k^{2}e_{1}P^{\nu} r_{\nu}=0,\label{t1}\\
\left(v{^{(1)}}_{1}\right)_{K}Z^{(1)}_{K}|_{\Xi=0}&=&-k^{2}e_{2} P^{\nu} r_{\nu}=0.\label{t2}
\end{eqnarray}
The substitution $\Xi=0$, guarantees that these constraints  will drop from the ramainder of the calculation. From the Eqs. (\ref{t1}) and (\ref{t2}), one can clearly assert that  there is only one new constraint given by
\begin{equation}
 \Theta=P^{\nu} r_{\nu}=0,\label{con3}
\end{equation}
because $e_{1}\neq0$ and $e_{2}\neq0$, by construction. It is worth remarking that the presence of this constraint is of essential importance for almost all the models proposed in the literature  for the description of the interaction between two relativistic particles. At the classical level, this constraint allows us to  eliminate the relative time variable, whereas at the quantum level this one allows us to remove some unphysical states. \cite{G1}. 

Now, another consistency condition must also be taken into account on (\ref{con3}) to obtain the following system:
\begin{equation}
f_{MJ}^{(2)}\dot{\xi}^{J}=Z_{M}^{(2)}(\xi)\label{equation3},
\end{equation}
with
\begin{equation}
f^{(2)}_{MJ}=
\left(
\begin{array}{cc} 
f^{(1)}_{KJ} \\ 
\frac{\delta\Theta}{\delta\xi^{J}}
\end{array}
\right)\hphantom{111}\&\hphantom{111}Z_{M}^{(2)}(\xi)=
\left(
\begin{array}{ccc} 
Z^{(1)}_{K} \\ 
0 \\
\end{array}
\right)\label{29},
\end{equation}
from which $f^{(1)}_{KJ}$ is given in Eq. (\ref{equation2}), while the matrix $\frac{\delta\Theta}{\delta\xi^{J}}$  takes the simple form, 
\begin{equation}
\label{eq30}
\frac{\delta\Theta}{\delta\xi^{J}}=
\small{
\left(
  \begin{array}{cccccc}
-P^{\mu}&r^{\mu}&P^{\mu}&r^{\mu}&0&0
 \end{array}
\right).}
\end{equation}
It is not difficult to see that this  matrix is singular. In fact, we get the following zero-modes
\begin{eqnarray}
\left(v^{(2)}{_{1}}\right)_{M}&=& \left(p_{1}^{\nu}, k^{2}r^{\nu}, 0, -k^{2}r^{\nu}, 0, 0, 1, 0,0 \right)\label{v4},\\
\left(v^{(2)}{_{2}}\right)_{M}&=& \left(0, k^{2}r^{\nu}, p_{2}^{\nu}, -k^{2}r^{\nu}, 0, 0, 0, 1,0 \right),\label{v3}\\
\left(v^{(2)}{_{3}}\right)_{M}&=& \left(r^{\nu},P^{\nu}, r^{\nu}, -P^{\nu}, 0, 0, 0, 0,1 \right).\label{v5}
\end{eqnarray} 
It is easy to verify that the contraction of the zero-modes (\ref{v4}) and (\ref{v3}) with $Z_{M}^{(2)}$ leads to an identity $\left(v^{(2)}{_{1,2}}\right)_{M}Z_{M}^{(2)}|_{\Xi, \Theta=0} = 0$, while the contraction of (\ref{v5}) with $Z_{M}^{(2)}$ leads to the following new constraint
\begin{equation}
\Omega=P_{\nu} \left(e_{1}p_{1}^{\nu}-e_{2}p_{2}^{\nu}\right)=0.\label{quar1}
\end{equation}
Before proceeding further,  it is interesting to note that if ones takes into account the explicit form of the constraints $\Phi_{1}$ and $\Phi_{2}$, the following relations can be deduced:
\begin{eqnarray}
2p_{1}^{\nu} P_{\nu}&=&P^{2}+\Delta+2\left(\Phi_{1}-\Phi_{2}\right),\\ 
2p_{2}^{\nu} P_{\nu}&=&P^{2}-\Delta-2\left(\Phi_{1}-\Phi_{2}\right), \label{relation}
\end{eqnarray}
here we have defined $\Delta\equiv m_{1}^{2}-m_{2}^{2}$. Accordingly,  inserting the two last expressions in (\ref{quar1}), we can rewrite the constraint (\ref{quar1}) in the following form:
\begin{equation}
\frac{1}{2}P^{2}\left[\left(e_{1}-e_{2}\right)+\frac{\Delta} {P^{2}}\left(e_{1}+e_{2}\right)\right]=0,
\end{equation}
where we have dropped a term proportional to $\left(\phi_{1}-\phi_{2}\right)$. Since we are interested  in the most general scenario with different rest masses $\Delta\neq 0$ and  $P^{2}\neq 0$ (massive case), it  is possible to replace the constraint (\ref{quar1}) by the following equivalent constraint:
\begin{equation}
\Gamma=\left(e_{1}-e_{2}\right)+\frac{\Delta}{P^{2}}\left(e_{1}+e_{2}\right)=0\label{quar2}.
\end{equation}
The motivation to replace $\Omega$ by $\Gamma$ is that the latter shows a relation between the einbeins $e_{1}$ and $e_{2}$.

Just as before, from time preservation of the constraint $\Gamma$ and the Eq. (\ref{equation3}) we inmediately obtain the folowing equation:
\begin{equation}
f_{NJ}^{(3)}\dot{\xi}^{J}=Z_{N}^{(3)}(\xi),\label{equation4}
\end{equation}
with
\begin{equation}
f^{(3)}_{NJ}=
\left(
\begin{array}{cc} 
f^{(2)}_{MJ} \\ 
\frac{\delta\Gamma}{\delta\xi^{J}}
\end{array}
\right) \hphantom{111}\&\hphantom{111}Z_{N}^{(3)}(\xi)=
\left(
\begin{array}{ccc} 
Z_{M}^{(2)} \\ 
0 \\
\end{array}
\right).
\end{equation}
Here the matrix $f_{MJ}^{(2)}$ has the same expression given by Eq. (\ref{29}), whereas $\frac{\partial\Gamma}{\partial\xi^{J}}$ becomes
\begin{equation}
\label{f3}
\frac{\partial\Gamma}{\partial\xi^{J}}=
\small{
\left(
  \begin{array}{cccccc}
0&-\Sigma P^{\mu}&0&-\Sigma P^{\mu}&\Lambda_{+}&-\Lambda_{-}
 \end{array}
\right),}
\end{equation}
where we have denoted: $\Lambda_{\pm}=1\pm\frac{\Delta}{P^{2}}$ and $\Sigma=2\frac{\Delta(e_{1}+ e_{2})}{(P^{2})^{2}}$. It is straightforward to show that, even after computing the linearly independent zero-modes that generate the kernel of the matrix $f_{NJ}^{(3)}$, no new constraints can be obtained. This means that our procedure is done, and thus  all the  obtained constraints form a complete and irreducible set. Once we have extracted all the physical constraints, we can define the extended Lagrangian, ${\mathcal{L}}{^{(E)}}$, adding the information of the constraints to the symplectic Lagrangian\footnote{ According to Faddeev-Jackiw approach, the constraints $\Phi$ must be incorporated into the canonical sector of the first-order Lagrangian, either as $\dot{\alpha}\Phi$ or $\alpha\dot{\Phi}$ with $\alpha$ some Lagrange multiplier. For practical reasons, it is better to replace $\lambda\rightarrow\dot{\alpha}$. The difference, being a total derivative, does not affect the equations of motion: $\lambda\Phi\rightarrow\lambda\Phi-\frac{d}{d\tau}(\alpha\Phi)=(\lambda-\dot{\alpha})\Phi-\alpha\dot{\Phi}$, and choosing $\lambda=\dot{\alpha}$.} (\ref{LS}), and evaluating the symplectic potential on the constraint surface. The extended Lagrangian reads
\begin{equation}
{\mathcal{L}}{^{(E)}} = p_{1\mu}\dot{x}_{1}^{\mu}+p_{2\mu}\dot{x}_{2}^{\mu}+\Phi_{1}\dot{\lambda}+\Phi_{2}\dot{\alpha}+\Theta\dot{\beta}+\Gamma\dot{\rho}.\label{NewL}
\end{equation}
In the above Lagrangian $\dot{\lambda}$, $\dot{\alpha}$, $\dot{\beta}$ and $\dot{\rho}$ are arbitrary  Lagrange multipliers asssociated to the constraints that enforce the stability of the constraints in time. Note that the symplectic potential has dropped  from the Lagrangian after the enforcement of  $\Phi_{1}, \Phi_{2},\Theta, \Gamma=0$, along with $V^{(E)}(\xi)=V|_{\Phi_{1}, \Phi_{2},\Theta, \Gamma=0}=0$. This fact shows the general covariance of the theory, and therefore, the dynamics will be clearly governed by the constraints. Thus, the Lagrangian (\ref{NewL}) contains all the necessary information to describe the interaction between the two massive relativistic particles. For the Lagrangian (\ref{NewL}), the new set of symplectic variables is 
\begin{equation}
{\xi}{^{(E)I}}= \left(x_{1}^{\mu}, p_{1}^{\mu},x_{2}^{\mu},p_{2}^{\mu},e_{1},e_{2},\lambda, \alpha, \beta,\rho\right)\label{var-final},
\end{equation}
whereas the corresponding one-form reads
\begin{equation}
a^{(E)}_{I}= (p_{1\mu}, 0, p_{2\mu}, 0,0,0,\Phi_{1},\Phi_{2},\Theta,\Gamma).
\end{equation}
 By using these symplectic variables, we can write the symplectic matrix, $f^{(E)}_{IJ}=\frac{\partial }{\partial\xi^{(E)I}}a^{(E)}_{J}-\frac{\partial }{\partial\xi^{(E)J}}a^{(E)}_{I}$, as follows:

\begin{eqnarray}
\label{eq}
f^{(E)}_{IJ}=
\small{
\left(
\begin{array}{cccccccccc}
 0   & -\eta^{\mu}_{\nu}  & 0  &0     &0&0& -k^{2}r_{\nu}&-k^{2}r_{\nu}& -P_{\nu}&0	 	 \\                                                                        
\eta^{\mu}_{\nu} & 0   & 0   & 0 &0&0&p_{1\nu}&0&r_{\nu}&-\Sigma P_{\nu} \\                                                                   
0   &0   &  0      & - \eta^{\mu}_{\nu}   & 0&0&k^{2}r_{\nu}&k^{2}r_{\nu}&P_{\nu}&0	\\
0   & 0   &  \eta^{\mu}_{\nu}    & 0  &0&0&0&p_{2\nu}&r_{\nu}&-\Sigma P_{\nu} \\
0&0&0&0&0&0&0&0&0&\Lambda_{+}\\
0&0&0&0&0&0&0&0&0&-\Lambda_{-}\\
k^{2}r^{\mu}&-p_{1}^{\mu}&-k^{2}r^{\mu}&0&0&0&0&0&0&0\\
k^{2}r^{\mu}&0&-k^{2}r^{\mu}&-p_{2}^{\mu}&0&0&0&0&0&0\\
P^{\mu}&-r^{\mu}&-P^{\mu}&-r^{\mu}&0&0&0&0&0&0\\
0&\Sigma P^{\mu}&0&\Sigma P^{\mu}&-\Lambda_{+}&\Lambda_{-}&0&0&0&0
 \end{array}
\right)}\label{matfin}.
\end{eqnarray}
It is interesting to see that the symplectic matrix (\ref{matfin}) remains degenerate, therefore the two-dimensional kernel is generated by the  following zero-modes:
\begin{eqnarray}
\left(v^{(E)}{_{1}}\right)_{I}&=& \left(\left(1-\frac{\Delta}{P^{2}}\right)p_{1}^{\nu} ,0,\left(1+\frac{\Delta}{P^{2}}\right)p_{2}^{\nu}, 0,0,0 ,-\left(1-\frac{\Delta}{P^{2}}\right),-\left(1+\frac{\Delta}{P^{2}}\right),0,0\right),\label{vf1} \\
\left(v^{(E)}{_{2}}\right)_{I}&=& \left(0, 0, 0,0,1 , \frac{P^{2}+\Delta}{P^{2}-\Delta}, 0, 0,0,0\right)\label{vf2}.
\end{eqnarray} 
However, we have shown that there are not further constraints in the system. Thus, the above observation implies that there might be a local gauge symmetry in the theory. 

In the following section, we will analyze the gauge symmetry throught the point of view of the symplectic approach,  which is encoded in the components of the remains zero-modes (\ref{vf1}) and (\ref{vf2}).

\section{Gauge symmetry}
In this section, we analyze the gauge symmetry for this model within the symplectic approach.  It was shown in \cite{barcelos, barcelos2, Montani1, Montani2}, that the degeneracy of the symplectic matrix (\ref{matfin}) and the fact that its remaining zero-modes are orthogonal to the gradient of the potential mean that these zero-modes generate locals displacements on the isopotential surface, in other words,  these zero-modes gives rise to the corresponding local gauge symmetries `$\delta_{G}$' on the constraints surface. This implies that we can identify the zero-modes with the generators of the gauge symmetry, that is
\begin{equation}
\delta_{G}\xi^{(E)}_{I}=\sum_{\alpha}v^{(E)}_{I\alpha}\epsilon^{\alpha},\label{generator}
\end{equation}
where $\{v_{\alpha}^{(E)}\}$ is the whole set of zero-modes of the degenerated symplectic matrix $f_{IJ}^{(E)}$, and $\epsilon\left(\tau\right)^{\alpha}$ stand for arbitrary gauge parameters defined on the configuration space.  According to (\ref{generator}), the full  gauge transformations for the symplectic variables $\xi^{(E)}_{I}$ that are generated by the zero-modes (\ref{vf1}) and (\ref{vf2}), are given by
\begin{eqnarray}
\delta_{G}x_{1}^{\mu}&=&\left(1-\frac{\Delta}{P^{2}}\right)p_{1}^{\mu}\epsilon_{1},\label{tra}\\
\delta_{G}p_{1}^{\mu}&=&0\label{tra1},\\
\delta_{G}x_{2}^{\mu}&=&\left(1+\frac{\Delta}{P^{2}}\right)p_{2}^{\mu}\epsilon_{1},\label{tra2}\\
\delta_{G}p_{2}^{\mu}&=&0,\label{tra3}\\
\delta_{G}e_{1}&=&{\epsilon}_{2},\label{tra4}\\
\delta_{G}e_{2}&=&\frac{P^{2}+\Delta}{P^{2}-\Delta}{\epsilon}_{2},\label{tra5}\\
\delta_{G}\lambda&=&-\left(1-\frac{\Delta}{P^{2}}\right)\epsilon_{1},\label{tra6}\\
\delta_{G}\alpha&=&-\left(1+\frac{\Delta}{P^{2}}\right)\epsilon_{1},\label{tra7}\\
\delta_{G}\beta&=&0,\label{tra8}\\
\delta_{G}\rho&=&0.\label{tra9}
\end{eqnarray}
Furthermore, using these gauge transformations, one can readily check that our extended Lagrangian  ${\mathcal{L}}{^{(E)}}$ (\ref{NewL}) transforms as
\begin{eqnarray}
\delta_{G}{\mathcal{L}}{^{(E)}}&=&\frac{d}{d\tau}\left[p_{1\mu}\delta_{G}x_{1}^{\mu}+p_{2\mu}\delta_{G}x_{2}^{\mu}+\Phi_{1}\delta_{G}\lambda+\Phi_{2}\delta_{G}\alpha\right]\nonumber\\
&&+\left(1-\frac{\Delta}{P^{2}}\right)\left[p_{1\mu}\left(\delta x_{1}^{\mu}\right)_{(E)}-\left(\delta\lambda\right)_{(E)}\right]\epsilon_{1}\nonumber\\
&&+\left(1+\frac{\Delta}{P^{2}}\right)\left[p_{2\mu}\left(\delta x_{2}^{\mu}\right)_{(E)}-\left(\delta\alpha\right)_{(E)}\right]\epsilon_{1},
\end{eqnarray}
where $\left(\delta x_{1}^{\mu}\right)_{(E)}$,  $\left(\delta x_{2}^{\mu}\right)_{(E)}$, $\left(\delta\alpha\right)_{(E)}$, $\left(\delta\lambda\right)_{(E)}$ are the equations of motion for the new set of symplectic variables $\xi^{(E)I}$,  which can be found in Appendix B. Hence, the action $S^{(E)}=\int d\tau{\mathcal{L}}{^{(E)}}$, is invariant on-shell up to a boundary term under the above gauge transformations generated by the zero-modes. As a consequence, the boundary term, that emerges from the gauge transformations  is
\begin{equation}
K=\epsilon_{1}\left[\left(1-\frac{\Delta}{P^2}\right)\left(p_{1}^{2}-\Phi_{1}\right)+\left(1+\frac{\Delta}{P^2}\right)\left(p_{2}^{2}-\Phi_{2}\right)\right] ,
\end{equation}
which demonstrate that, effectively, the set of all gauge transformations for the symplectic variables, listed in (\ref{tra})-(\ref{tra9}), are symmetries of the extended Lagrangian (\ref{NewL}) generated by the zero-modes (\ref{vf1}) and (\ref{vf2}). Classically, this implies a conserved charge, however, the associated Noether's charge is zero, because the gauge conserved quantities are always zero.

On the other hand, it is worth remarking that even although Eqs. (\ref{tra})-(\ref{tra9}) correspond to the fundamental gauge symmetry of the theory, these transformations are not diffeomorphisms `$\delta_{\mathrm{Diff}}$' ( reparametrization symmetry ). Nevertheless, we  can redefine the gauge parameters as
\begin{equation}
\epsilon_{1}\equiv\left(\frac{P^{2}}{P^{2}-\Delta}\right)\epsilon e_{1}\hphantom{111}\&\hphantom{111}\epsilon_{2}\equiv\frac{d}{d\tau}(\epsilon e_{1}),\label{lie}
\end{equation}
where $\epsilon\left(\tau\right)$ denote an arbitrary function that parametrizes the two massive relativistic particles.  In this manner from the fundamental gauge transformations (\ref{tra})-(\ref{tra9}) and the mapping (\ref{lie}), we obtain
\begin{eqnarray}
\delta_{\mathrm{Diff}}x_{1}^{\mu}&=&\epsilon e_{1}p_{1}^{\mu},\label{gauge1}\\
\delta_{\mathrm{Diff}}x_{2}^{\mu}&=&\epsilon e_{2}p_{2}^{\mu}+\left(\frac{P^{2}}{P^{2}-\Delta}\right)\epsilon\Gamma,\\\label{gauge2}
\delta_{\mathrm{Diff}}e_{1}&=&\frac{d}{d\tau}(\epsilon e_{1}),\\\label{gauge3}
\delta_{\mathrm{Diff}}e_{2}&=&\frac{d}{d\tau}(\epsilon e_{2})+\left(\frac{P^{2}}{P^{2}-\Delta}\right)\frac{d}{d\tau}\epsilon\Gamma,\label{gauge4}
\end{eqnarray}
which are diffeomorphisms modulo terms proportional to the constraints. It should be noticed that, the individual diffeomorphism invariance of the two free relativistic particles is broken explicitly due to the interaction between the two particles. Finally, let's quickly check that (\ref{gauge1})-(\ref{gauge4}) are indeed a symmetry of the Lagrangian ${\mathcal{L}}$, and find the boundary term:
\begin{eqnarray}
\delta_{\mathrm{Diff}}{\mathcal{L}}&=&\frac{d}{d\tau}\left[\epsilon{\mathcal{L}}\right]+\frac{d}{d\tau}\left[\epsilon p_{1\mu}\left(\delta p_{1}^{\mu}\right)+\epsilon p_{2\mu}\left(\delta p_{2}^{\mu}\right)\right]+\epsilon e_{1}\left[p_{1\mu}\frac{d}{d\tau}\left(\delta x_{1}^{\mu}\right)+\frac{d}{d\tau}\left(\delta e_{1}\right)\right]\\\nonumber
&&+\epsilon e_{2}\left[p_{2\mu}\frac{d}{d\tau}\left(\delta x_{2}^{\mu}\right)+\frac{d}{d\tau}\left(\delta e_{2}\right)\right],
\end{eqnarray}
which turns out be precisely a boundary term, but modulo terms proportional to the equations of motion: $\left(\delta x_{1}^{\mu}\right)$, $\left(\delta x_{2}^{\mu}\right)$, $\left(\delta p_{1}^{\mu}\right)$, $\left(\delta p_{2}^{\mu}\right)$,  $\left(\delta e_{1}\right)$ and $\left(\delta e_{2}\right)$, which are defined in (\ref{eom1})-(\ref{eom6}). 

On the other hand, for the case of equal rest masses $\Delta=0$, the physical constraint $\Gamma$ in (\ref{quar2}) takes the following form: $\bar{\Gamma}=(e_{1}-e_{2})=0$, because $P^{2}\neq0$ (the condition is guaranteed by the conservation of the total momentum $P^{\mu}$). At this stage, the gauge transformations can be determined from the expressions (\ref{gauge1})-(\ref{gauge4}) by  taking $\Delta\rightarrow0$. Therefore, the gauge transformations for the case of equal rest masses turn out to be
\begin{eqnarray}
\delta_{\mathrm{Diff}}x_{1}^{\mu}&=&\epsilon e_{1}p_{1}^{\mu},\label{gauge1.1}\\
\delta_{\mathrm{Diff}}x_{2}^{\mu}&=&\epsilon e_{2}p_{2}^{\mu}+\epsilon\bar\Gamma,\\\label{gauge2.1}
\delta_{\mathrm{Diff}}e_{1}&=&\frac{d}{d\tau}(\epsilon e_{1}),\\\label{gauge3.1}
\delta_{\mathrm{Diff}}e_{2}&=&\frac{d}{d\tau}(\epsilon e_{2})+\frac{d}{d\tau}\epsilon\bar\Gamma.\label{gauge4.1}
\end{eqnarray}
Taking into account the form (\ref{momentum}) of the canonical momentum and the fact that $\bar\Gamma=0$, these transformations now read
\begin{eqnarray}
\delta_{\mathrm{Diff}}x_{1}^{\mu}&=&\epsilon \dot{x}_{1}^{\mu},\label{gauge1.2}\\
\delta_{\mathrm{Diff}}x_{2}^{\mu}&=&\epsilon \dot{x}_{2}^{\mu},\\\label{gauge2.2}
\delta_{\mathrm{Diff}}e_{1}&=&\frac{d}{d\tau}(\epsilon e_{1}),\\\label{gauge3.2}
\delta_{\mathrm{Diff}}e_{2}&=&\frac{d}{d\tau}(\epsilon e_{2}).\label{gauge4.2}
\end{eqnarray}
These gauge transformations are equivalent to the gauge symmetries in (\ref{gauge1.1})-(\ref{gauge4.1}). In conclusion,  the diffeomorphisms symmetry is not an independent symmetry. In fact, the diffeomorphisms can be obtained from the fundamental gauge symmetries (\ref{tra})-(\ref{tra9}) by mapping the gauge parameters. In addition, the generators of such gauge transformations have been identified directly from the structure of the zero-modes, thereby making evident that the zero-modes of the symplectic two-form matrix encode all the information of the gauge symmetry of this model.
\section{Gauge fixing and quantization bracket}
Finally,  to define the quantization brackets and also find the physical degrees of freedom, one can fix the gauge freedom associated to the singularity in the symplectic matrix (\ref{matfin}) by imposing additional conditions (gauge conditions). Due to the gauge invariance (\ref{gauge1})-(\ref{gauge4}),  we can choose the following conditions (the conformal gauge) $e_{1}=1$ ($\Upsilon_{1}=e_{1}-1=0$),  $e_{2}=1$ ($\Upsilon_{2}=e_{2}-1=0$), $\beta=const$ ($\dot{\beta}=0$) and $\rho=const$ ($\dot{\rho}=0$), and  add these gauge fixing conditions to extended Lagrangian (\ref{NewL}), which can be rewriten as
\begin{equation}
{\mathcal{L}}{^{(F)}} = p_{1\mu}\dot{x}_{1}^{\mu}+p_{2\mu}\dot{x}_{2}^{\mu}+\Phi_{1}\dot{\lambda}+\Phi_{2}\dot{\alpha}+\left(\Theta-\gamma\right)\dot{\beta}+\left(\Gamma-\delta\right)\dot{\rho}+\Upsilon_{1}\dot{\mu}+\Upsilon_{2}\dot{\tau},\label{fg}
\end{equation}
where the  new arbitrary Lagrange multipliers enforcing the gauge conditions are $\dot{\mu}$,   $\dot{\tau}$, $\dot{\gamma}$ and $\dot{\delta}$. Then, from the above Lagrangian one obtains the following set of symplectic variables
\begin{equation}
{\xi}{^{(F)I}}= \left(x_{1}^{\mu}, p_{1}^{\mu},x_{2}^{\mu},p_{2}^{\mu},e_{1},e_{2},\lambda, \alpha, \beta,\rho, \mu, \tau, \gamma, \delta\right),\label{Fj}
\end{equation}
with the new one-form given by
\begin{equation}
a^{(F)}_{I}= (p_{1\mu}, 0, p_{2\mu}, 0,0,0,\Phi_{1},\Phi_{2},\Theta-\gamma,\Gamma-\delta, \Upsilon_{1}, \Upsilon_{2},0,0).
\end{equation}
Using these symplectic variables, we find that the corresponding symplectic matrix $f^{(F)}_{IJ}=\frac{\partial }{\partial\xi^{(F)I}}a^{(F)}_{J}-\frac{\partial }{\partial\xi^{(F)J}}a^{(F)}_{I}$ , in the conformal gauge, is given by
\begin{eqnarray}
\label{Fg}
f^{(F)}_{{IJ}}=
\tiny{
\left(
\begin{array}{cccccccccccccc}
 0   & -\eta^{\mu}_{\nu}  & 0  &0     &0&0& -k^{2}r_{\nu}&-k^{2}r_{\nu}& -P_{\nu}&0&0&0&0&0	 	 \\                                                                        
\eta^{\mu}_{\nu} & 0   & 0   & 0 &0&0&p_{1\nu}&0&r_{\nu}&-\Sigma P_{\nu}&0&0&0&0 \\                                                                   
0   &0   &  0      &  -\eta^{\mu}_{\nu}   & 0&0&k^{2}r_{\nu}&k^{2}r_{\nu}&P_{\nu}&0&0&0&0&0	\\
0   & 0   &  \eta^{\mu}_{\nu}    & 0  &0&0&0&p_{2\nu}&r_{\nu}&-\Sigma P_{\nu}  &0&0&0&0\\
0&0&0&0&0&0&0&0&0&\Lambda_{+}&1&0&0&0\\
0&0&0&0&0&0&0&0&0&-\Lambda_{-}&0&1&0&0\\
k^{2}r^{\mu}&-p_{1}^{\mu}&-k^{2}r^{\mu}&0&0&0&0&0&0&0&0&0&0&0\\
k^{2}r^{\mu}&0&-k^{2}r^{\mu}&-p_{2}^{\mu}&0&0&0&0&0&0&0&0&0&0\\
P^{\mu}&-r^{\mu}&-P^{\mu}&-r^{\mu}&0&0&0&0&0&0&0&0&1&0\\
0&\Sigma P^{\mu}&0&\Sigma P^{\mu}&-\Lambda_{+}&\Lambda_{-}&0&0&0&0&0&0&0&1\\
0&0&0&0&-1&0&0&0&0&0&0&0&0&0\\
0&0&0&0&0&-1&0&0&0&0&0&0&0&0\\
0&0&0&0&0&0&0&0&-1&0&0&0&0&0\\
0&0&0&0&0&0&0&0&0&-1&0&0&0&0\\
 \end{array}
\right),}
\end{eqnarray}
 We can observe that $f^{(F)}_{IJ}$ is non-singular, i.e., no zero-modes exist.  Therefore, we compute its inverse $\left(f^{(F)}_{IJ}\right)^{-1}$, which is  given by
\begin{equation}
\label{Fg}
\left(
\tiny{
\begin{array}{cccccccccccccc}
 0   & \eta^{\nu}_{\mu}-\frac{p_{1\mu}P^{\nu}}{P^{2}} & -\Lambda_{2}p_{1\mu}r^{\nu}  &\frac{p_{1\mu} P^{\nu}}{P^{2}}   &0&0& 0&\Lambda_{1}r_{\mu}& 0&0&0&0&-\Lambda_{\mu}^{\nu}r_{\nu}&-\Sigma P_{\mu}	 	 \\                                                                        
-\eta^{\nu}_{\mu}+\frac{p_{1\mu}P^{\nu}}{P^{2}} & 0   & -\frac{p_{2\mu}P^{\nu}}{P^{2}}   & 0 &0&0&-\frac{P_{\mu}}{P^{2}}&\frac{P_{\mu}}{P^{2}}&0&0&0&0&0&0 \\                                                                   
\Lambda_{2}p_{1\mu}r^{\nu}    &\frac{p_{2\mu}P^{\nu}}{P^{2}}  &  0      &\eta^{\nu}_{\mu}-  \frac{p_{2\mu}P^{\nu}}{P^{2}}   & 0&0&-\Lambda_{2}r_{\mu}&0&0&0&0&0&-\Lambda_{\mu}^{\nu}r_{\nu}&-\Sigma P_{\mu}	\\
-\frac{p_{1\mu}P^{\nu}}{P^{2}}   & 0   & -\eta^{\nu}_{\mu} +\frac{p_{2\mu}P^{\nu}}{P^{2}}    & 0  &0&0&\frac{P_{\mu}}{P^{2}}&-\frac{P_{\mu}}{P^{2}}&0&0&0&0&0&0\\
0&0&0&0&0&0&0&0&0&0&-1&0&0&0\\
0&0&0&0&0&0&0&0&0&0&0&-1&0&0\\
0&\frac{P^{\nu}}{P^{2}}&\Lambda_{2}r^{\nu}&-\frac{P^{\nu}}{P^{2}}&0&0&0&-\Lambda P^{\nu}r_{\mu}&0&0&0&0&\Lambda_{2}{P_{\nu}r^{\nu}}&0\\
-\Lambda_{1}r^{\nu}&-\frac{P^{\nu}}{P^{2}}&0&\frac{P^{\nu}}{P^{2}}&0&0&\Lambda P^{\nu}r_{\mu}&0&0&0&0&0&\Lambda_{1}{P_{\nu}r^{\nu}}&0\\
0&0&0&0&0&0&0&0&0&0&0&0&-1&0\\
0&0&0&0&0&0&0&0&0&0&0&0&0&-1\\
0&0&0&0&1&0&0&0&0&0&0&0&0&\Lambda_{+}\\
0&0&0&0&0&1&0&0&0&0&0&0&0&-\Lambda_{-}\\
\Lambda_{\mu}^{\nu}r^{\mu}&0&\Lambda_{\mu}^{\nu}r^{\mu}&0&0&0&-\Lambda_{2}{P^{\nu}r_{\nu}}&-\Lambda_{1}{P^{\nu}r_{\nu}}&1&0&0&0&0&0\\
\Sigma P^{\nu}&0&\Sigma P^{\nu}&0&0&0&0&0&0&1&-\Lambda_{+}&\Lambda_{-}&0&0
 \end{array}}
\right),
\end{equation}
here  $\Lambda=\frac{1}{k^{2}r^{2}P^{2}}$, $\Lambda_{\mu}^{\nu}=\Lambda_{2}p_{1\mu}P^{\nu}-\eta_{\mu}^{\nu}$ , $\Lambda_{2}=\frac{p_{2}\cdot P}{k^{2}r^{2}P^{2}}$,  $\Lambda_{1}=\frac{p_{1}\cdot P}{k^{2}r^{2}P^{2}}$. 

Thus, the generalized Faddeev-Jackiw bracket $\{,\}_{F-J}$ for any two elements of the symplectic variables set (\ref{Fj}) is defined by:
\begin{equation}
\{\xi_{I}^{(F)}(x),\xi_{J}^{(F)}(y)\}_{F-J}\equiv\left(f_{IJ}^{(F)}\right)^{-1}.
\end{equation}
In this way, the Feddeev-Jackiw brackets between basic variables are
\begin{eqnarray}
\{x_{1}^{\nu}, p_{1\mu}\}_{F-J} &=& \eta^{\nu}_{\mu}-\frac{p_{1\mu}P^{\nu}}{P^{2}},	\\
\{x_{1}^{\nu}, x_{2\mu}\}_{F-J} &=& -\frac{p_{2}\cdot P}{k^{2}r^{2}P^{2}}p_{1\mu}r^{\nu}, 	\\
\{x_{1}^{\nu}, p_{2\mu}\}_{F-J} &=& \frac{p_{1\mu}P^{\nu}}{P^{2}},	\\
\{x_{2}^{\nu},p_{1\mu}\}_{F-J} &=& \frac{p_{2\mu}P^{\nu}}{P^{2}}, \\
\{x_{2}^{\nu}, p_{2\mu}\}_{F-J} &=&\eta^{\nu}_{\mu}- \frac{p_{2\mu}P^{\nu}}{P^{2}},
\end{eqnarray}
while the remaining brackets vanish. Note that the position space has a non-commutative structure as a result of the interaction.  On the other hand, it is clear that the limit $P^{2}\rightarrow 0$ of this algebra is singular, and therefore, the limit $P^{2}\rightarrow0$ of the model of two massive relativistic particles interacting  itself does not exist. The canonical quantization can be made by the replacement of the Faddeev-Jackiw brackets by the operator commutation relations according to
\begin{equation}
\{A,B\}_{F-J}\longrightarrow \frac{1}{i\hbar}\left[\hat{A},\hat{B}\right](\mathrm{Commutator}).
\end{equation}
In addition, we can carry out the counting of degrees of freedom as follows: starting with $18$ canonical variables $(e_{1}, e_{2}, x_{1}^{\mu}, x_{2}^{\mu}, p_{1}^{\mu}, p_{2}^{\mu})$, we end up with $8$ independent constraints $(\Phi_{1},\Phi_{2},\Upsilon_{1},\Upsilon_{2}, \Theta,\Gamma,\beta,\rho)$ after imposing the gauge-fixing term. Thus, we conclude that this model has $12$ physical degree of freedom, as claimed in \cite{dominici}.
\section{Conclusions}
In the present paper, we have used the Faddeev-Jackiw symplectic approach to analyze the dynamical structure of two interacting relativistic particles \cite{dominici}, which can also be interpreted as a bigravity model in one dimension. One of the interesting features of this analysis is the fact that the dynamical content of this model has been extracted by studying only the properties of the symplectic matrix and its corresponding zero-modes. In particular,  we have determined the structure of all the physical constraints directly from the zero-modes of the corresponding symplectic matrix through the contraction of these zero-modes with the gradient of the symplectic potential. Furthermore, we have inferred the full fundamental gauge transformations in terms of the remaining zero-modes, which means that these are indeed the generators of the local gauge symmetry under which all physical quantities are invariant. We have identified also the time reparametrization (Diffeomorphisms) symmetry  by mapping the gauge parameters appropriately. We have also shown that the Lagrangian is invariant under the symmetries generated by the zero-modes, and also under the diffeomorphisms symmetry. Finally, it has been shown that the conformal gauge condition on the Lagrangian (\ref{NewL}) renders a non-degenerate symplectic matrix $f_{IJ}^{(F)}$, whose inverse allows one to identify  the quantization brackets. On the other hand, we believe that it would be interesting to perform the symplectic analysis, especially for bigravity theories, written in the first-order formalism. An analysis of this problem, at least in the case of the 2 + 1 dimensional model of the  bigravity action proposed in \cite{bigravity}, will be published elsewhere.
\newline
\newline
\textbf{Acknowledgements}\\[1ex]
This work has been supported by CONACyT under grant number CB-2014/237503. The author  thanks  G. Tavares-Velasco for reading a draft version of this paper and alerting to various typos.
\appendix
\section{Equation of motion}
In this appendix, we indicate how the equation of motion can be obtained in the first-oder formalism. Let us start with a first-order a Lagrangian which is first order in time derivatives, i.e.,
\begin{equation}
{\mathcal{L}}(\xi, \dot{\xi})=a_{I}(\xi)\dot{\xi}^{I}-V(\xi).
\end{equation}
Then, the associated equation of motion  can be obtained by performing the first variation of the Lagrangian:
\begin{eqnarray}
\delta{\mathcal{L}}(\xi, \dot{\xi})&=&\delta\xi^{J}\frac{\partial a^{I}}{\partial \xi_{J}}\dot{\xi}_{I}+a^{I}{}\delta\dot{\xi}_{I}-\delta\xi^{J}\frac{\partial}{\partial\xi_{J}} V(\xi)\\
&=&\delta\xi^{J}\frac{\partial a^{I}}{\partial \xi_{J}}\dot{\xi}_{I}+\frac{d}{dt}\left(a^{I}{}\delta{\xi}_{I}\right)-\dot{a}^{I}\delta\dot{\xi}_{I}-\delta\xi^{J}\frac{\partial}{\partial\xi_{J}} V(\xi)\\
&=&\delta\xi^{J}\frac{\partial a^{I}}{\partial \xi_{J}}\dot{\xi}_{I}-\dot{\xi}^{J}\frac{\partial a^{I}}{\partial \xi_{J}}\delta{\xi}_{I}-\delta\xi^{J}\frac{\partial}{\partial\xi_{J}} V(\xi)\\
&=&\delta\xi^{J}\left[\left(\frac{\partial}{\partial\xi_{J}}a^{I}-\frac{\partial}{\partial\xi_{I}}a^{J}\right)\dot{\xi}_{I}-\left(\frac{\partial}{\partial\xi_{J}}V\right)\right].
\end{eqnarray}
Therefore, if $\delta\xi^{J}$ are independent, the equations of motion can be written as
\begin{equation}
f_{IJ}\dot{\xi}^{J}-\frac{\partial }{\partial\xi_{I}}V(\xi)=0\label{a1},
\end{equation}
where $f_{IJ}$ is defined by
\begin{equation}
f_{IJ}\equiv \frac{\partial}{\partial\xi_{I}}a^{J}-\frac{\partial}{\partial\xi_{J}}a^{I}.
\end{equation}
If $f_{IJ}$ is nonsingular, it is possibe to write Eq. (\ref{a1}) as
\begin{equation}
\dot{\xi}_{I}=\left (f_{IJ}\right)^{-1}\frac{\partial}{\partial\xi_{J}}V.
\end{equation}
\section{Equations of motion associated to ${\mathcal{L}}{^{(E)}}$}
In this appendix we present the equations of motion associated to the extended Lagrangian ${\mathcal{L}}{^{(E)}}$ (\ref{NewL}). Using the equations of motion expression (\ref{a1}), one can compute the equations of motion in the form:
\begin{equation}
f_{IJ}^{(E)}\xi^{(E)J}=0\label{a2},
\end{equation}
where the set of symplectic variables is given by ${\xi}{^{(E)I}}= \left(x_{1}^{\mu}, p_{1}^{\mu},x_{2}^{\mu},p_{2}^{\mu},e_{1},e_{2},\lambda, \alpha, \beta,\rho\right)$ and the symplectic matrix $f_{IJ}^{(E)}$ is defined in (\ref{vf2}). Then, from Eq. (\ref{a2}) it is easy to obtain the explicit form of  equations of motion:
\begin{eqnarray}
(\delta x_{1}^{\mu})_{(E)}&:&-\dot{p}_{1}^{\mu}-k^{2}r^{\mu}\left(\dot{\lambda}+\dot{\alpha}\right)-P^{\mu}\dot{\beta}=0,\\
(\delta p_{1}^{\mu})_{(E)}&:&\dot{x}_{1}^{\mu}+p_{1}^{\mu}\dot{\lambda}+r^{\mu}\dot{\beta}-\Sigma P^{\mu}\dot{\rho}=0,\\
(\delta x_{2}^{\mu})_{(E)}&:&-\dot{p}_{2}^{\mu}+k^{2}r^{\mu}\left(\dot{\lambda}+\dot{\alpha}\right)+P^{\mu}\dot{\beta}=0,\\
(\delta p_{2}^{\mu})_{(E)}&:&\dot{x}_{2}^{\mu}+p_{2}^{\mu}\dot{\lambda}+r^{\mu}\dot{\beta}-\Sigma P^{\mu}\dot{\rho}=0,\\
(\delta e_{1}^{\mu})_{(E)}&:&\Delta_{+}\dot{\rho}=0,\\
(\delta e_{2}^{\mu})_{(E)}&:&-\Delta_{-}\dot{\rho}=0,\\
(\delta \lambda)_{(E)}&:&-\dot{\phi}_{1}=0,\\
(\delta \alpha)_{(E)}&:&-\dot{\phi}_{2}=0,\\
(\delta \beta)_{(E)}&:&-\dot{\Theta}=0,\\
(\delta \rho)_{(E)}&:&-\frac{1}{2}\frac{d}{d\tau}{P}^{2}-\dot{\Gamma}=0.
\end{eqnarray}

\end{document}